\documentclass[preprint, showpacs,preprintnumbers,amsmath,amssymb,nofootinbib]{revtex4}
\usepackage{amssymb}
\usepackage{amsfonts}

\usepackage{epsf}
\usepackage{graphicx}
\usepackage{subfigure}

\usepackage{float}
\begin{document}
	\newcommand{\be}[1]{\begin{equation}\label{#1}}
	\newcommand{\ee}{\end{equation}}
	\newcommand{\bea}{\begin{eqnarray}}
	\newcommand{\eea}{\end{eqnarray}}
	\newcommand{\bed}{\begin{displaymath}}
	\newcommand{\eed}{\end{displaymath}}
	\def\disp{\displaystyle}
	
	\def\gsim{ \lower .75ex \hbox{$\sim$} \llap{\raise .27ex \hbox{$>$}} }
	\def\lsim{ \lower .75ex \hbox{$\sim$} \llap{\raise .27ex \hbox{$<$}} }
	
	\title{Prospects of strongly lensed repeating fast radio bursts: complementary constraints on dark energy evolution}
	
	\author{Bin Liu, Zhengxiang Li$^{}\footnote{Electronic address: zxli918@bnu.edu.cn}$, Zong-Hong Zhu}
	
	\address{Department of Astronomy, Beijing Normal University, Beijing 100875, China}

	\begin{abstract}
		Fast radio bursts (FRBs) are highly dispersed and probably extragalactic radio flashes with millisecond-duration. Recently, the Canadian Hydrogen Intensity Mapping Experiment (using the CHIME/FRB instrument) has reported detections of 13 FRBs during a pre-commissioning phase. It is more exciting that one of the 13 FRBs is a second source of repeaters which suggests that CHIME/FRB and other wide-field sensitive radio telescopes will find a substantial population of repeating FRBs. We have proposed strongly lensed repeating FRBs as a precision cosmological probe, e.g. constraining the Hubble constant and model-independently estimating the cosmic curvature. Here, we study complementary constraints on the equation of state of dark energy from strongly lensed FRBs to currently available popular probes. It is found that, in the framework of Chevalier-Polarski-Linder parametrization, adding time delay measurement of 30 strongly lensed FRB systems to cosmic microwave background radiation and type Ia supernovae can improve the dark energy figure of merit by a factor 2. In the precision cosmology era, this improvement is of great significance for studying the nature of dark energy. 
	\end{abstract}

	\maketitle
	\renewcommand{\baselinestretch}{1.5}

	\section{Introduction}
	Fast radio bursts (FRBs) are mysterious radio transients with milli-second duration and excess dispersion measure (DM) with respect to the Galactic value~\cite{lorimer07,thornton13,petroff15,petroff16,katz18}. The localization of FRB 121102, which is the first repeating source occurring at a dwarf star-forming galaxy at $z=0.19$, has confirmed the cosmological origin~\cite{scholz16,spitler16,chatterjee17,marcote17,tendulkar17}. Since these bright flashes originate from cosmological distances, their observed DMs which characterize a line-of-sight integral of the free electron density on the way to Earth should be mainly produced by the intergalactic medium (IGM). In this case, it was found that FRBs appear to be at distances up to Gpc and have cosmological redshift $z$ of 0.5 to 1 by modeling the relation of average dispersion with respect to distance~\cite{ioka03, inoue04, zheng14, deng14}. Because of this merit, FRBs have been proposed as promising cosmological and astrophysical probes, such as cosmological implications from DM and redshift $z$ measurements~\cite{zhang14, deng14, gao14, zhou14, walters18}, locating the ``missing" baryons~\cite{mcquinn14}, tracing the linear large-scale structure of the universe~\cite{masui15}, tests for the Einstein's equivalence principle~\cite{wei15, nusser16, tingay16}, constraints on rest mass of the photon~\cite{wu16,shao17}, magnetic fields in the IGM~\cite{akahori16}, and cosmic proper distance measurement~\cite{yu}
	
	This phenomenon was first discovered at radio frequencies $\sim1.4$ GHz and so far has been observed at as high as 8 GHz in one case~\cite{gajjar18}. Recently, although during a pre-commissioning phase when the sensitivity and field-of-view were not yet at design specifications, the Canadian Hydrogen Intensity Mapping Experiment (CHIME) has detected 13 FRBs at frequencies as low as 400 MHz with the CHIME/FRB instrument~\cite{chime19a}. Moreover, it is breathtaking that a second source of repeaters (FRB 180814.J0422+73), one of these 13 detections, has been reported~\cite{chime19b}. This second repeater suggests that CHIME/FRB and other wide-field sensitive radio telescopes will find a considerable number of repeating FRBs. Since the estimation for the all-sky FRB rate is high~\cite{thornton13,champion16} and FRBs can be visible out to cosmological distances $z\gtrsim1$~\cite{zhang18}, a substantial number of (repeating) FRBs are expected to be gravitationally lensed by intervening galaxies~\cite{hilbert08,li14,wagner18}.Roughly speaking, a few$\times 10^3-10^4$ FRBs are expected to be detected per year by CHIME/FRB~\cite{CHIME/FRB}. If repeating ones are $\sim$ $1/10$ of FRBs as observed and the typical probability of galaxy-galaxy strong lensing would be $10^{-4}-10^{-3}$, it is likely for CHIME/FRB having dozens of strongly lensed repeaters in a few $\times 10$ years.With prospects of some other currently avaliable and forthcoming large-scale radio surveys, including ASKAP~\cite{bannister18}, UTMOST~\cite{caleb16}, and HIRAX~\cite{newburgh16}, it is probly to accumulate dozens of strongly lensed repeaters in a shorter period and thus these interseting systems are worthy of a consideration. For instance, gravitationally lensed FRBs have been proposed as a powerful probe of compact dark matter~\cite{wang18,munoz16} and the motion of the source~\cite{dai17}. More recently, we have shown that time delay difference measurements between multiple images of strongly lensed repeating FRBs might be able to estimate the Hubble constant at sub-percent precision~\cite{li18}. This result will be very helpful to arbitrate the intractable tension between the Hubble constant directly estimated from local distance ladder and that constrained from the cosmic microwave background observations in the standard $\Lambda$ plus cold dark matter ($\Lambda$CDM) model~\cite{riess16,planck15}.
	
	In cosmology, the nature of late-time accelerated expansion is a deep mystery. An exotic energy content with negative pressure, named dark energy, is usually introduced to account for this unexpected phenomenon. In the past two decades, a great deal of efforts via various programs of popular cosmic probes, such as type Ia supernovae (SNe Ia), baryon acoustic oscillations (BAOs), and cosmic microwave background (CMB) radiation, have been carried out to understand the nature of dark energy. One of the most effective way for grasping properties of dark energy is to study the equation of state (EOS),  and it is usually parametrized as the most popular Chevalier-Polarski-Linder (CPL) form~\cite{chevalier01, linder03}, $w(z)=p(z)/\rho(z)=w_0+w_a*z/(1+z)$. In this route, complementary constraints from different kind of observations are of utmost importance due to cross-checking the systematic bias and breaking the degeneracy between parameters $w_0$ and $w_a$. Strong gravitational lensing is a robust tool for  
	both astrophysics and cosmology~\cite{treu10}. It was pointed out that time delay measurements of strongly lensed quasar systems can provide complementarity with other prevalent cosmological probes to constrain cosmological parameters~\cite{linder04}. The time delay distance, a combination of three angular diameter distances, is derived from time delay measurements and lens profile modeling and has different degeneracy directions from those of other popular probes such as SNe Ia and CMB. Specifically, in the most popular CPL parametrization, $w_0$ and $w_a$ have positive correlation instead of anticorrelation in other probes when lenses lie in the redshift rang $0-0.6$. It was suggested that the dark energy figure of merit can be improved by a factor of 5 by adding 150 well-measured lensed quasar systems, i.e. ``golden lens", to the combination of SNe Ia and CMB observations available at that time~\cite{linder11}. Recently, we have proposed strongly lensed repeating FRBs as a precision cosmological probe due to, compared with traditional lensed quasars, significant enhancements in the precision of time delay measurements and lens modeling~\cite{li18}. Here, we investigate complementary constraints on dark energy evolution from strongly lensed FRBs to currently available popular probes and find that adding time delay measurements of 30 such systems to CMB plus SNe Ia can improve the constraining power on dark energy by a factor 2. At the age of precision cosmology era, this improvement is of great importance for exploring the nature of dark energy.

	\section{Methodology}\label{method}
	
	In a gravitational lensing system, the difference in arrival time between image $i$ and image $j$, (time delay, $\Delta t_{i,j}$), can be predicted via~\cite{treu10,treu16}
	\begin{equation}
	\Delta t_{i,j}=\frac{(1+z_{l})D_{\Delta t}}{c}\Delta \phi_{i,j},
	\end{equation}
	where $z_l$ is the redshift of lens and $c$ is the light speed. The time delay distance $D_{\Delta t}$ is relative to the following three angular diameter distances,  
	\begin{equation}
	D_{\Delta t}=\frac{D_l(z_l)D_s(z_s)}{D_{ls}(z_l,z_s)},
	\end{equation}
	where $z_s$ is the redshift of source. $D_l$, $D_s$, and $D_{ls}$ are angular diameter distances from the lens to the observer, from the source to the observer, and from the lens to the source, respectively. $\Delta \phi_{i,j}$ is the difference of Fermat potential of $i$, $j$ images and it can be written as
	\begin{equation}
	\Delta \phi_{i,j}=\frac{(\theta_{i}-\beta)^2}{2}-\Psi(\theta_{i})-\frac{(\theta_{j}-\beta)^2}{2}+\Psi(\theta_{j}),
	\end{equation}
	where $\theta_{i}$ ,$\theta_{j}$ are angular positions of $i$, $j$ images and $\beta$ is the angular position of the source. $\Psi$ is the two-dimensional lens potential. For the traditional lensed quasars, one has to monitor light curves for a long time to measure time delay between images. For lensed FRBs, the time delay between images can be measured to extremely high precision because of the large ratio $\sim10^9$ between the typical galaxy-lensing delay time $\sim\mathcal{O}$(10 days) and the narrow widths of bursts $\sim\mathcal{O}$(ms). 
	
	In the flat Friedmann-Lema{\^i}tre-Robertson-Walker (FLRW)  framework, angular diameter distance can be expressed as,
	\begin{equation}
	D(z;\mathbf{p})=\frac{1}{1+z}\frac{c}{H_0}\int_0^z \frac{dz'}{E(z';\mathbf{p})},
	\end{equation}
	where $H_0$ is the Hubble constant, $E(z;\mathbf{p})=H(z)/H_0$ is the dimensionless expansion rate, $\textbf{p}$ is a set of cosmological parameters.
	When we take the CPL parametrization to characterize the dark energy evolution with respect to redshift, where
	\begin{equation}
	w(z)=w_0+w_a \frac{z}{1+z},
	\end{equation}
	the expansion rate can be written as
	\begin{equation}
	E^2(z;\mathbf{p})=\Omega_m(1+z)^3+(1-\Omega_m)(1+z)^{3(1+w_0+w_a)}\exp\bigg(-\frac{3w_az}{1+z}\bigg).
	\end{equation}
	In this case, it was found that there is a positive correlation between $w_0$ and $w_a$ in the time delay distance in the redshift range $0<z_l<0.6$~\cite{linder04}. This correlation direction is orthogonal to some other popular probes where $w_0$ and $w_a$ are negatively correlated. Therefore, strong lensing time delay is an ideal complementary tool to some currently popular probes for studying dark energy evolution.

	From the observation of lensing events, on one hand, we can measure the time delay between lensed signals. On the other hand, we can identify the host galaxy of the source and obtain high-resolution images. These observations are essential to get precise and accuracy information about the Fermat potential of the lens. Then, the time delay distance is derived and can be subsequently applied to infer cosmological information via the likelihood function $L \sim e^{-\chi^2/2}$, 
	\begin{equation}
	\chi^2= \sum_i\bigg[\frac{D_{\Delta t}^{th}(z_l, z_s; \mathbf{p})-D_{\Delta t,i}^{obs}}{\sigma_{D_{\Delta t,i}}^{obs}}\bigg]^2.
	\end{equation}
	
	At present, sources of almost all strong lensing systems are quasars and observations of time delay are mainly in optical band and only a small number of lensing systems are well measured. For time delay measurements by monitoring light curves of quasars, the average relative uncertainty is about $3\%$ shown by the first time delay challenge (TDC1)~\cite{liao15}. The average relative uncertainty of Fermat potential difference or lens modeling is also about $3\%$ due to contaminations of dazzling AGNs in the center of the source~\cite{suyu17}. Fortunately, future detections of lensed FRBs will significantly improve the present situation. First, the time delay in strongly lensed FRB systems can be accurately determined owing to the small ratio between the short duration of FRB signals, $\sim \mathcal{O}(10^{-3}~s)$, and the typical galaxy-lensing delay time $\sim\mathcal{O}$(10 days). Second, high quality images of the host galaxy can be obtained since there is no central dazzling AGN. This advantage will improve the precision of lens modeling by a factor of $\sim4$~\cite{li18}. That is, the relative error of Fermat potential reconstruction can be reduced to $\sim 0.8\%$.It should be pointed out that this result was obtained on the basis of the assumption that qualities of host galaxy images obtained from the \textit{Hubble Space Telescope} or the near future \textit{James Webb Space Telescope} are high enough to distinguish among different lens models and the lens mass profiles could be correctly reconstructed. Here, we study the complementarity of this promising probe on investigating the dark energy evolution. Here, we study the complementarity of this promising probe on investigating the dark energy evolution.

	\section{Results}\label{results}
	In order to estimate the constraining power of future lensed FRBs on dark energy evolution, we should address two issues: the redshift distribution of strongly lensed FRB systems and the corresponding uncertainty level of time delay distance determination. The later one consists of the following three ingredients: the uncertainty of time delay measurement, the uncertainty of Fermat potential difference, and the extra uncertainty from the mass distributed along the line of sight.
	
\begin{figure}
		\centering
		\includegraphics[scale=0.68]{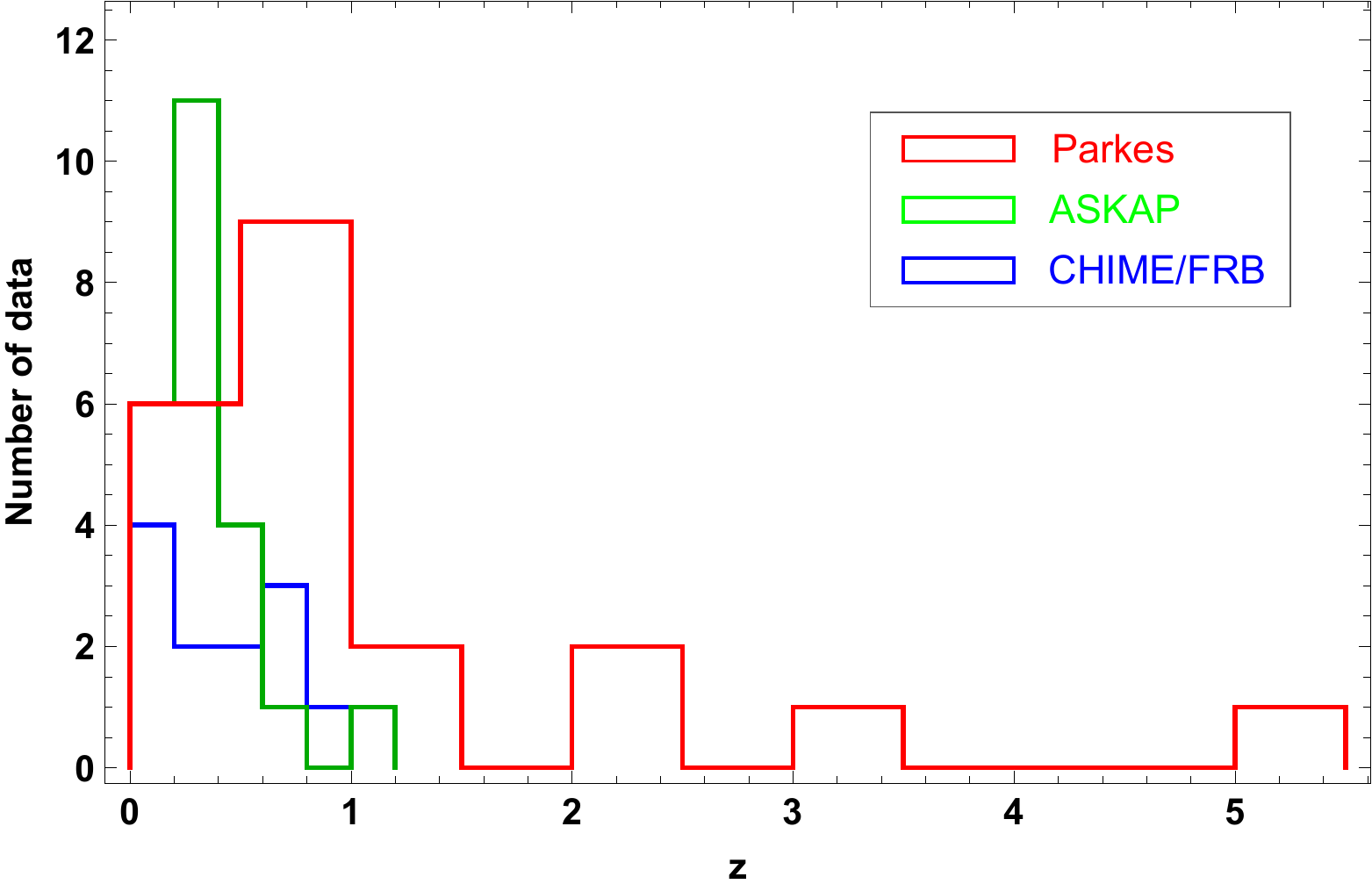}
		\caption{Redshift distribution of FRBs detected by different instruments.}
		\label{redshift}
	\end{figure}

	For redshifts of currently available FRBs, different from previous estimation using a simple relataion between DM and $z$ proposed by Ioka~\cite{ioka03}, we re-estimate them with a more precise DM-$z$ relation given in~\cite{deng14}. It is found that the inferred $z$ values are systematically greater than previously estimated ones with several FRBs having $z >1$. Inferred redshifts for currently available FRBs from different facilities are shown in Figure~\ref{redshift}. We consider two possible distributions suggested in~\cite{munoz16}. The first one invokes a constant comoving number density, so that the number of FRBs in a shell of width $dz$ at redshift $z$ is proportional to the comoving volume of the shell $dV(z)$~\cite{oppermann16}. By introducing a Gaussian cutoff at some redshift $z_{\rm{cut}}$ to represent an instrumental signal-to-noise threshold, the constant-density distribution function $N_{\rm{const}}(z)$ is expressed as
	\begin{equation}
	N_{\rm{const}}(z)=\mathcal{N}_{\rm{const}}\frac{\tilde{\chi}^2(z)}{H(z)(1+z)}e^{-{D^{\rm{L}}}^2(z)/[2{D^{\rm{L}}}^2(z_{\rm cut})]},
	\end{equation}
	where $\tilde{\chi}(z)$ is the comoving distance and $D^{\rm{L}}$ is the luminosity distance. $\mathcal{N}_{\rm{const}}$ is a normalization factor to ensure that the integration of  $N_{\rm{const}}(z)$ is unity and $H(z)$ is the Hubble parameter at redshift $z$. The second distribution requires that FRBs follow the star-formation history (SFH)~\cite{caleb16}, so that
	\begin{equation}
	N_{\rm{SFH}}(z)=\mathcal{N}_{\rm{SFH}}\frac{\dot{\rho}_*(z)\tilde{\chi}^2(z)}{H(z)(1+z)}e^{-{D^{\rm{L}}}^2(z)/[2{D^{\rm{L}}}^2(z_{\rm cut})]},
	\end{equation}
	where $\mathcal{N}_{\rm{SFH}}$ is the normalization factor and is chosen to have $N_{\rm{SFH}}(z)$ integrated to unity. The density of star-formation history is parametrized as 
	\begin{equation}
	\dot{\rho}_*(z)=h\frac{\alpha+\beta z}{1+(z/\gamma)^\delta},
	\end{equation}
	with $\alpha=0.017,~\beta=0.13,~\gamma=3.3,~\delta=5.3$, and $h=0.7$ \cite{cole01,hopkins06}. For these two FRB distribution functions, a cutoff $z_{\rm cut}=1$ is chosen to match redshifts of currently detected events. In our previous analysis~\cite{li18}, we have found that these two distribution functions do not lead to significant difference in cosmological implications. Therefore, $N_{\rm const}$ with a cutoff $z_{\rm cut}=1$ is used in our following analysis. With the redshift of the source in hand, we next calculate the lensing probability for a source locating at redshift $z_s$~\cite{schneider92}: 
	\begin{equation}\label{lenprob}
	P=\int_0^{z_s}dz_l\frac{dD_p}{dz_l}\int_0^{\infty}dMn(M,z_l)\sigma(M,z_l).
	\end{equation}
	$D_p$ is the proper distance between the observer and the lens, $n(M,z_l)$ is the proper number density at of dark matter halos in the mass range $M$ and $M+dM$ at $z_l$, $\sigma(M,z_l)$ is the lensing cross-section with mass $M$. The mass of the halos is assumed in the range $10^{10}h^{-1}M_{\odot}<M<2\times10^{13}h^{-1}M_{\odot}$ and the mass profile is assumed as the simple SIS model~\cite{li02,li14}. For a source at a given redshift ${z_s}$, the lensing probability for a lens with mass in the range $10^{10}h^{-1}M_{\odot}<M<2\times10^{13}h^{-1}M_{\odot}$ at redshift $z_l$ can be derived from Equation \ref{lenprob}.  For a given $z_s$, there is a specific $z_l$ where strong lensing is most likely to happen (see Figure 2 in \cite{li14}). We take these $z_l$ as the lens redshifts in our following simulations. 
	
	For the corresponding uncertainty levels in determining the time delay distance, the first ingredient, time delay measurement, can be considered to be very accurate and the $\delta \Delta t$ can be taken as 0 because of the short duration of FRBs. Meanwhile, the measure of Fermat potential will also be significantly improved. For repeating FRBs, we can localize them with very long baseline interferometers (VLBI) and find their host galaxy. Unlike images of traditional lensed quasars where there is a very bright AGN locating in the center making it is very difficult to determine the Fermat potential accurately, images of the host galaxy of repeaters are free of this contamination and the uncertainty of Fermat potential difference estimation can reach to $0.8\%$ \cite{li18}. In addition, we also consider the intractable systematic error caused by the mass distribution along the line of sight. According to state-of-the-art level, we take it as $2\%$~\cite{cristian17, tihhonova18, bonvin17}. After considering these uncertainties, we can determine the uncertainty of the time delay distance measurement of strongly lensed FRB systems.  
	
	In order to investigate the complementarity of strongly lensed FRB systems to already existing popular probes on constraining dark energy EOS, we include the latest $Planck$ 2018 CMB observations~\cite{planck18,lu18} and SNe Ia from the upcoming DES Hybrid 10-field Survey(Table 14 of ~\cite{bernstein12}). The parameters set is $\{\Omega_m,H_0,w_0,w_a\}$. We use the minimization function to get the parameters sets of minimum $\chi^2$ and run the simulation 10000 times with different random seeds.
	
\begin{figure}
		\centering
		\includegraphics[scale=0.98]{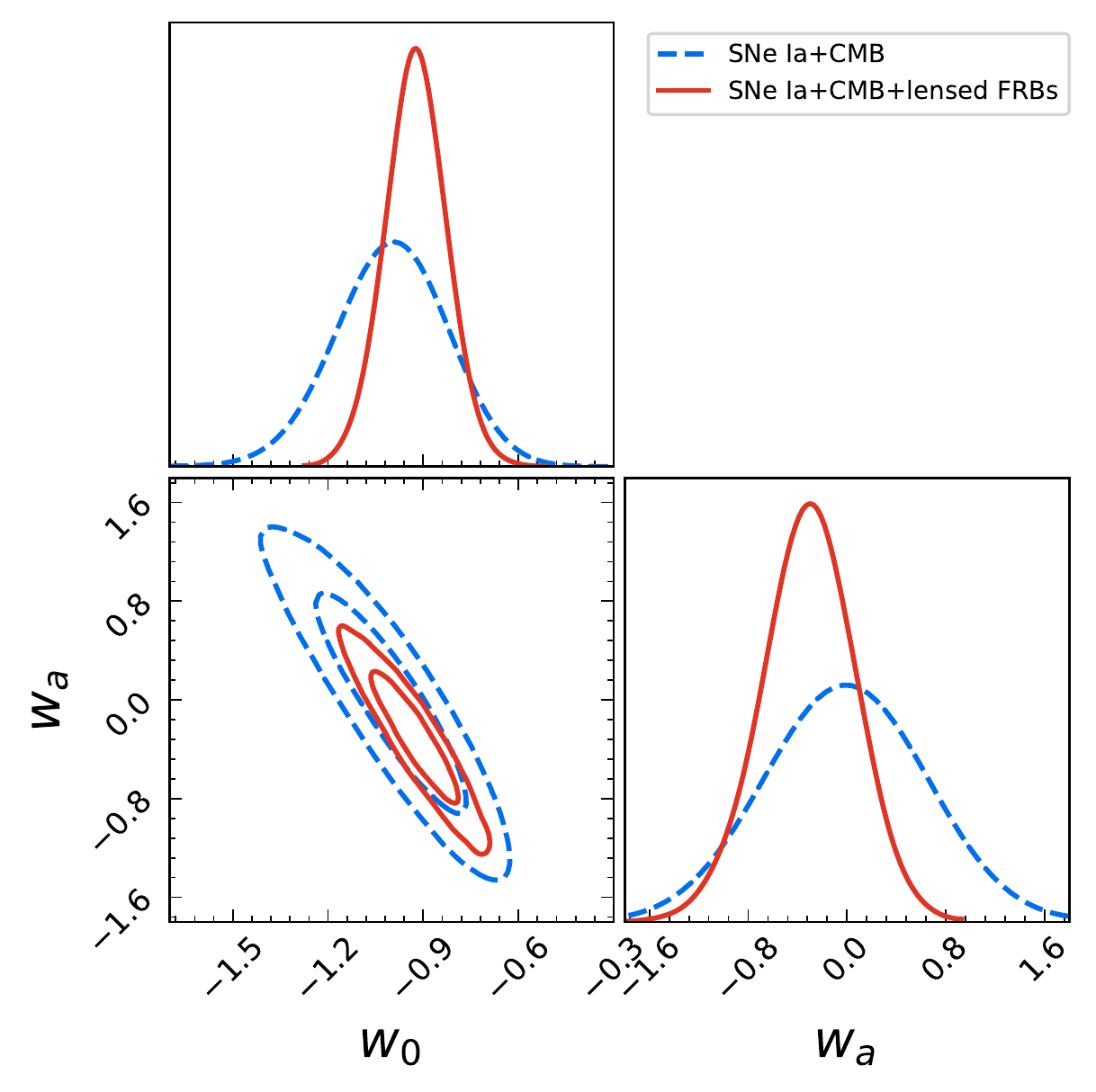}
		\caption{Marginalized PDFs and the $68\%,~95\%$ confidence contours of the dark energy parameters $w_0$ and $w_a$, 30 lensed FRB systems plus SNe Ia and CMB are considered.}
		\label{30_events}
	\end{figure}

\begin{figure}
		\centering
		\includegraphics[scale=0.98]{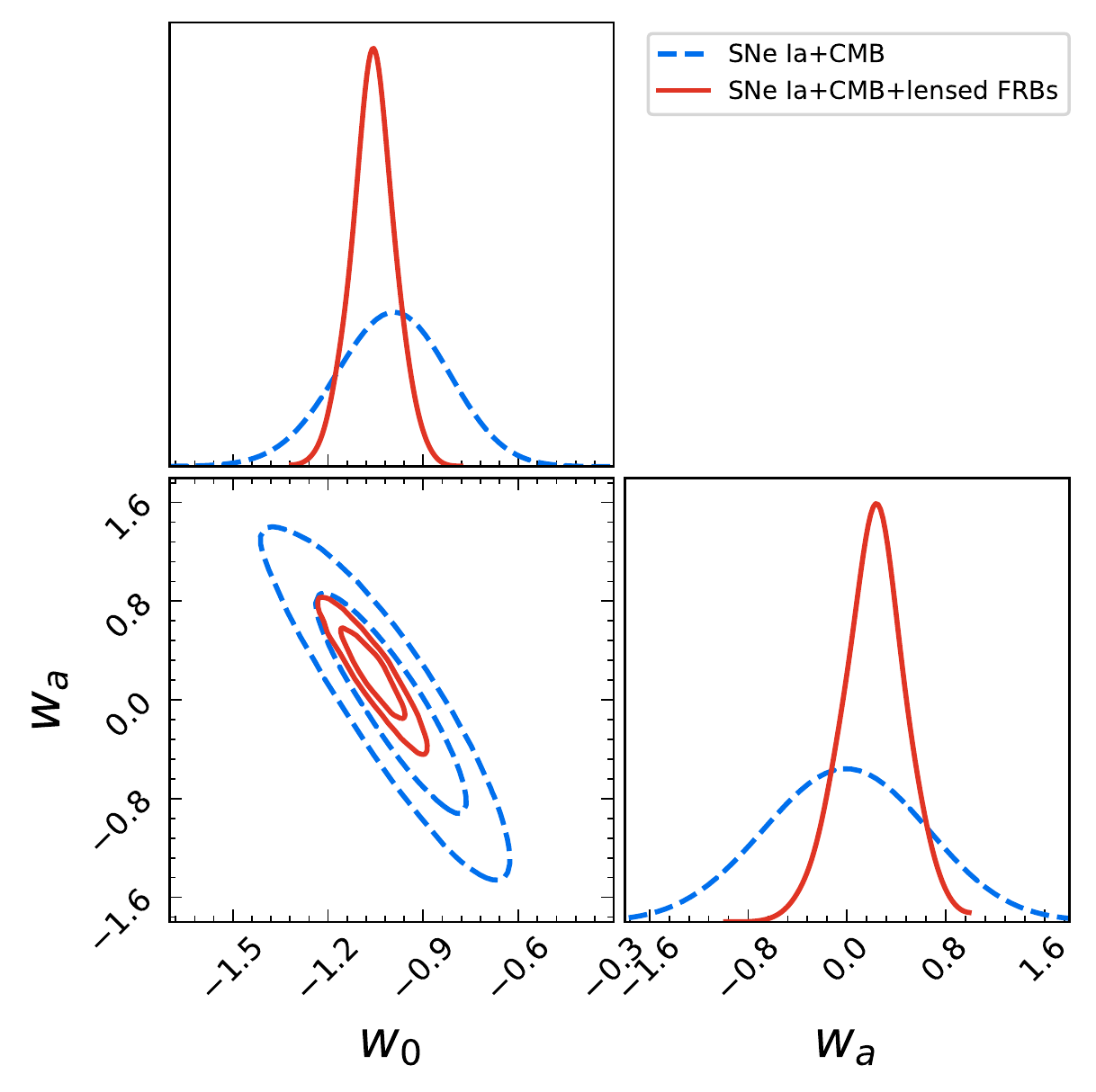}
		\caption{Marginalized PDFs and the $68\%,~95\%$ confidence contours of the dark energy parameters $w_0$ and $w_a$, 100 lensed FRB systems plus SNe Ia and CMB are considered.}
		\label{100_events}
	\end{figure}

\begin{figure}
		\centering
		\includegraphics[scale=0.50]{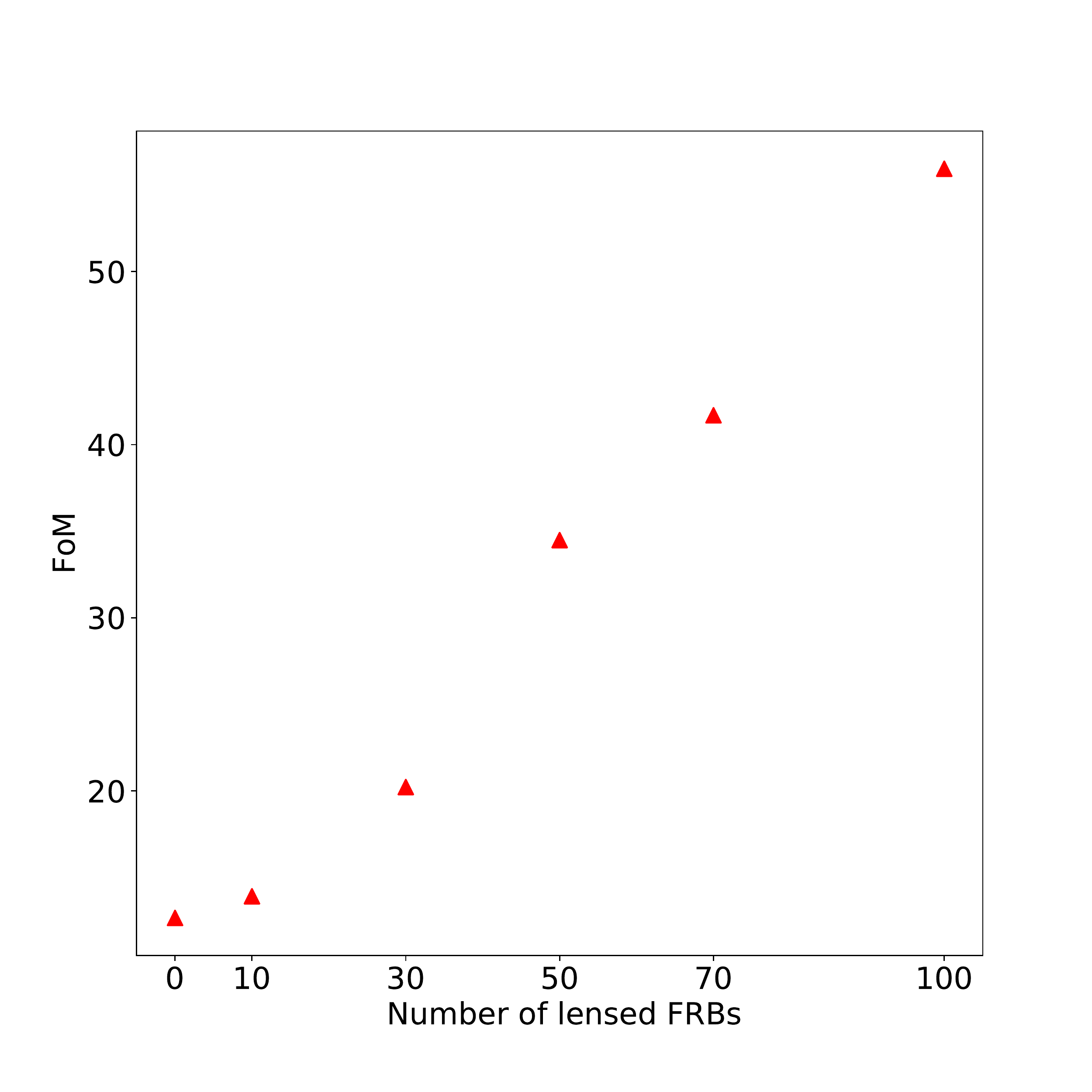}
		\caption{The figure of merit (FoM) of constraint on the dark energy equation of state from currently available popular probes (SNe Ia+CMB) plus different number of time delay measurement of strongly lensed FRB systems.}
		\label{FoM}
	\end{figure}

	We consider several tens of lensed FRB events which are likely to be collected by upcoming wide-field sensitive radio telescopes to estimate the complementary constraints on $w_0$ and $w_a$. Results are shown in Figures (\ref{30_events}, \ref{100_events}) and Table \ref{tab1}. In order to quantify the improvement of constraints on dark energy equation of state after taking lensed FRB systems into account, we apply the figure of merit (FoM)~\cite{albrecht06,wang08,dossett11,sendra11} which is proportional to the inverse area of the error ellipse in the $w_0-w_a$ plane, 
	\begin{equation}
	\mathrm{FoM}=[\mathrm{det}C(w_0, w_a)]^{-1/2}, 
	\end{equation}
	where $C(w_0, w_a)$ is the covariance matrix of $w_0-w_a$ after marginalizing over all other cosmological parameters. Larger FoM implies stronger constraint on the parameters since it relates to a smaller error ellipse. FoMs of constraint on the dark energy equation of state from currently available $Planck$ 2018 CMB observations and upcoming DES SNe Ia together with different number of time delay measurement of strongly lensed FRB systems are plotted in Figure \ref{FoM}. It is suggested that the constraining power is improved by a factor of 2 when 30 strongly lensed FRB systems are considered to combine with other popular probes. In the era of precision cosmology, constraints from observations of almost all popular probes are consistently in the favor of the standard $\Lambda$CDM model. However, the well-known irreconcilable Hubble constant tension might strongly indicates the evidence of dark energy evolution. Therefore, in this sense, improvement (by a factor of 2) of constraints on the dark energy equation of state is utmost importance for exploring the nature of dark energy.
	
	\begin{table}
		\begin{center}{\scriptsize
				\renewcommand\arraystretch{1.8}
				\begin{tabular}{p{6cm}p{4cm}p{0.75cm}}
					\hline \hline 
					number of lensed FRBs    & $\sigma_{w_0}$  & $\sigma_{w_a}$ \\
					\hline
					0 (only SNe Ia+CMB)                         & 0.148 & 0.563 \\
					\hline
					10                          & 0.117 & 0.441 \\
					\hline
					30                          & 0.093 & 0.360 \\
					\hline
					50                          & 0.082  & 0.312 \\
					\hline
					70                          & 0.076 & 0.292\\
					\hline
					100                         & 0.069 & 0.255 \\
					\hline \hline 
				\end{tabular}}
				\caption{The $1\sigma$ uncertainties of the dark energy parameters for CPL parameterization constrained from different number of lensed FRBs considered.}\label{tab1}
			\end{center}
		\end{table}

		\section{Conclusions and Discussions}\label{Cons}
		Strong gravitational lensing has been so far widely used in astrophysics and cosmology. In the early days, cosmological implications were first obtained from statistical distribution of observations and theoretical images separation or using the lensing effect of galaxy clusters~\cite{paczynski81,sereno02,meneghetti05,gilmore09,jullo10}. Later, the distance ratios derived from observed velocity dispersion of the gravitational lens and images separation were proposed to study cosmology~\cite{biesiada06,grillo08,biesiada10,cao12,cao15}. Moreover, the distance ratio from strong lensing observations together with other distance measurements can also be used as an independent tool to test FLRW metric and constrain the curvature of the universe~\cite{rasanen15,li18b}. In this distance ratio method, a relative simple profile, i.e. singular isothermal spherical or singular isothermal elliptical profile, is usually used to characterize the lens mass distribution of all measured systems. This treatment might lead to non-negligible systematic uncertainties~\cite{li18b}. In addition to the above-mentioned methods, time delay measurements of strong gravitational lensing systems where each lens is individually modeled and therefore systematical bias can be significantly decreased have been gradually applied for cosmological investigations. Especially, in the most popular CPL scenario, time delay is an effective cosmological probe with different $w_0,~w_a$ degeneracy direction from those of other currently popular probes. This virtue is very helpful to precisely constrain the dark energy evolution. However, for traditional lensed quasar systems, the precisions for lens profile modeling and time delay measurements have been limited since there is a dazzling AGN in the center of the source and the timescale of light-curve variation of the source is comparable with the scale of time delay between images. 
		
		Fast radio bursts are a relatively new astrophysical mystery and the field of FRB science is currently thriving. The second source of repeating FRBs implies that sensitive wide-field radio telescopes, such as CHIME/FRB, will detect a substantial number of repeaters. The observation of lensed repeating FRBs will also be possible in the future and these lensed systems can be used as cosmological probes. Compared to conventional lensed quasars, lensed FRB systems have overwhelming advantages in time delay measurement and lens profile modeling due to the high time-resolution of these transients and no dazzling AGN in the center of the source.  In this paper, we quantify the complementarity of time delay of upcoming strongly lensed FRB systems to currently popular probes, i.e. SNe Ia and CMB. It is found that the dark energy figure of merit can be improved by a factor 2 by combining SNe Ia and CMB with time delay of 30 strongly lensed FRB systems. In the era of precision cosmology, any improvement will give a great helpfulness for studying the nature of dark energy or even understanding the physical mechanism of cosmic acceleration.  Here, it should be clarified that this
conclusion is inferred with relatively optimistic estimations for event rate and error analysis. However, considering overwhelming advantages of these interesting systems and their potential impact on fundamental physics, large-scale surveys and follow-up facilities, together with great efforts in improving lens modeling will make this expectation
likely to achieve in the near future.
		\section{Acknowledgements}
		This work was supported by the National Natural Science Foundation of China under Grants Nos. 11505008, and 11633001, the Strategic Priority Research Program of the Chinese Academy of Sciences, Grant No. XDB23040100.

	\end{document}